\documentclass[a4paper,12pt,openright]{article}
\usepackage{color,theorem}
\usepackage[T1]{fontenc}
\usepackage{a4,amsmath,amssymb,time}
\usepackage{graphicx}
\usepackage{booktabs}
\usepackage{array} 
\usepackage{hyperref}

\date{\today\ \now}

\definecolor{darkgreen}{rgb}{0.06,0.39,0.13}
\definecolor{darkblue}{rgb}{0.06,0.39,0.63}

{\begin{list}{\normalsize\textcolor{red}{$\bullet$}}{\setlength{\leftmargin}{0.2in}\setlength{\itemsep}{0.5ex}}}
{\end{list}\normalsize}

\newenvironment{list1}%
{\begin{list}{\normalsize\textcolor{red}{$\bullet$}}{\setlength{\leftmargin}{0.2in}\setlength{\itemsep}{0.5ex}}}
{\end{list}\normalsize}

{\begin{list}{\normalsize\textcolor{red}{$\bullet$}}{\setlength{\leftmargin}{0.2in}\setlength{\itemsep}{0.5ex}}}
{\end{list}\normalsize}

\newcommand{\ontopof}[2]{\genfrac{}{}{0pt}{}{#1}{#2}}

\def\tightplus{\medmuskip=1.5mu\relax}
\def\hypk#1#2#3#4{\mathop{F{}}\mathchoice{\tightplus
  \hbox{$\displaystyle\biggl({\ontopof{#2}{#3}}\Big\vert\,{#4}\!\biggr)$}%
  \lower\fontdimen11\mathsym\hbox{$\scriptstyle\!#1$}}%
 {\bigl({\ontopof{#2}{#3}}\vert\mskip2mu#4\bigr)\lower\fontdimen12\mathsym
   \hbox{$\scriptstyle\!#1$}}%
 {}{}}  

\title{\textcolor{blue}{Asymptotic expansions for enumerating connected labelled graphs}}

\author{Keith Briggs\\
  {\tt Keith.Briggs@bt.com}\\
  BT Research\\
  Suffolk IP5 3RE, UK
}
\def\today{\number\year\space
 \ifcase\month\or January\or February\or March\or April\or May\or June\or
   July\or August\or September\or October\or November\or December\fi
     \space\number\day
}
\date{\today\ \ \now}

\begin{document}

\maketitle


\begin{center}
Abstract
\end{center}
\begin{center}
 {\fbox{%
  \begin{minipage}{13.5cm}
    I compute several terms of the asymptotic expansion of the
    number of connected labelled graphs with $n$ nodes and $m$ edges, for
    small $k=m-n$.  I thus identify an error in a recent paper of Flajolet \emph{et al}.
  \end{minipage}
 }}
\end{center}



\section{Introduction}

Consider the problem of computing the number $c(n,m)$ of connected labelled
graphs with $n$ nodes and $m=n-1,n,n+1,\dots$ edges, for fixed small $m$ as
$n\rightarrow\infty$  ([slo03], sequence A057500, and links therein).  From
this, we can compute the probability of a randomly chosen labelled graph being
connected, which is useful in various applications to communication networks.

In a recent paper Flajolet \emph{et al}. [fss04], the authors used ingenious
analytic methods to compute that the number of labelled connected graphs with
$n$ nodes and excess $=k\geqslant 2$ (excess is defined as the number of edges
minus the number of nodes) is asymptotically
$$
A_k(1)\sqrt{\pi}\left(\frac{n}{e}\right)^n \left(\frac{n}{2}\right)^{\frac{3k-1}{2}}\left[\frac{1}{\Gamma(3k/2)}+\frac{A'_k(1)/A_k(1)-k}{\Gamma((3k-1)/2)}\sqrt{\frac{2}{n}}+{{\cal{O}}\left(\frac{1}{n}\right)}\right]
$$
where 
$A_k(1)$ is given in terms of Airy functions; the first few values being as in
table~\ref{table1}.

\begin{table}[h]\tiny
\begin{center}
\begin{tabular}{rrrrrrrr}
\toprule
$k$&$1$&$2$&$3$&$4$&$5$&$6$&$7$\\
\midrule
$A_k(1)$ &$5/24$& $5/16$& $1105/1152$& $565/128$& $82825/3072$& $19675/96$& $1282031525/688128$\\
$A'_k(1)$&$19/24$& $65/48$& $1945/384$& $21295/768$& $603965/3072$& $10454075/6144$& $1705122725/98304$\\
\bottomrule
\end{tabular}
\caption{Coefficients in the formula of Flajolet \emph{et al}.}
\label{table1}
\end{center}
\end{table}

However, I found that this result agreed poorly in comparisons with exact
counts, which can be easily computed from the known generating functions.  I
did comparisons with exact counts for up to $n=1000$ nodes and for excess
$k=2,3,\dots,8$.   The results led me to suspect that the factor $(n/e)^n$
should be simply $n^n$, and that the second term in the square brackets should
have a minus sign.   The purpose if this note is therefore to compute more
terms of the asymptotic expansions (by different methods) to confirm these
suspicions.  The first term in these asymptotic expansions was already known
thanks to a recurrence relation (involving an implicitly defined quantity) due
to Bender \emph{et al}. [bcm90].

The asymptotic expansion of the number $c(n,n+k)$ of labelled connected
graphs with $n$ nodes and excess $k\geqslant -1$ has the form
$$
c(n,n+k)\sim n^{n+(3k-1)/2}\;\sum_{j=0}^{\infty}\; a_j(k)\,\xi^{1-((k+j)\;\text{mod}\;2)}\,n^{-j/2}
$$
where $\xi\equiv\sqrt{2\pi}$, and the coefficients $a_j(k)$ are rational.  This
structure in fact allows surprisingly reliable estimates of $a_j(k)$ for $j+k$
less than about 5 simply by fitting least-squares polynomials in $x=n^{-1/2}$
to the exact data.  By this means I obtained the estimates shown in
table~\ref{est}.

\begin{table} 
\begin{center}
\begin{tabular}{rcrrrrrr}
\toprule
$k$  & type & $[n^0]$   & $[n^{-1/2}]$ & $[n^{-1}]$ & $[n^{-3/2}]$& $[n^{-2}]$ & $[n^{-5/2}]$\\
\midrule
$ 0$ & unicycle & ${\xi\frac{1}{4}}$   & ${-\frac{7}{6}}$ & ${\xi\frac{1}{48}}$ &  ${\frac{131}{270}}$ & ${\xi\frac{1}{1152}}$ & ${-\frac{4}{2835} ?}$\\
$ 1$ & bicycle & ${\frac{5}{24}}$   &  ${-\xi\frac{7}{24}}$      &  ${\frac{25}{36}}$ &${-\xi\frac{7}{288}}$    &     ${-\frac{79}{3240}?}$&\\
$ 2$ & tricycle & ${\xi\frac{5}{256}}$ & ${-\frac{35}{144}}$  &   ${\xi\frac{1559}{9216}}$   &  {$-\frac{55}{144}$}   &   &     \\
$ 3$ & quadricycle & ${\frac{221}{24192}}$  & ${-\xi\frac{35}{1536}}$ & &   &  &\\
\bottomrule
\end{tabular}
\end{center}
\caption{Coefficients in the series for $c(n,n+k)/n^{n+(3k-1)/2}$, conjectured from numerical experiments.  Here and elsewhere $[n^x]f(n)$ means the coefficient of $n^x$ in $f(n)$.}
\label{est}
\end{table}

\section{Generating functions}

The exponential generating function (egf) for labelled graphs is 
$$g(w,z)=\sum_{n=0}^{\infty}\; (1+w)^{n \choose 2} z^n/n!.$$
This means that
$n!\,[w^m z^n]\,g(w,z)$
is the number of labelled graphs with $m$ edges and $n$ nodes.
The exponential generating function for all connected labelled graphs
is therefore
\begin{eqnarray}
c(w,z)&=&\log(g(w,z))\nonumber\\ &=&z+w\frac{z^2}{2}+(3w^2+w^3)\frac{z^3}{6}+(16w^3+15w^4+6w^5+w^6)\frac{z^4}{4!}+\dots\nonumber.
\end{eqnarray}
I will now compute the asymptotic expansion of $c(n,n+k)/n^{n+\frac{3k-1}{2}}$.
This needs some preliminary manipulations concerning the quantities $Q$, $W$
and $t$.


\begin{table}
\begin{center}
\begin{tabular}{rrrrrr}
\toprule
$k$  & type & B: $[n^0]$ & F: $[n^0]$ & F: $[n^{-1/2}]$ & $[n^{-1}]$\\
\midrule
$ 0$ & unicycle & ${\xi\frac{1}{4}}$   & &  & \\ 
$ 1$ & bicycle & ${\frac{5}{24}}$   & &  & \\ 
$ 2$ & tricycle & ${\xi\frac{5}{256}}$ & ${\xi\frac{5}{256}}$& ${\frac{35}{144}}$  &  \\
$ 3$ & quadricycle & ${\frac{221}{24192}}$ & ${\frac{221}{24192}}$ & ${\xi}\frac{35}{1536}$  &    \\
$ 4$ & pentacycle & ${\xi\frac{113}{196608}}$   & ${\xi\frac{113}{196608}}$  &  ${\frac{221}{20736}}$&  \\
\bottomrule
\end{tabular}
\end{center}
\caption{
Some (possibly incorrect) coefficients for $c(n,n+k)/n^{n+(3k-1)/2}$ from the literature.
B: from [bcm90]; F: from [fss04] (with removal of factor $e$).
Here $\xi\equiv\sqrt{2\pi}$.   
Note that $c(n,n-1)/n^{n-2}=1$.
Missing values are not available in the literature. 
}
\end{table}

\subsection{$Q$}

The results in this section follow from theory available in [jklp93] and [fgkp95].  Ramanujan's $Q$-function [ram11] is defined for $n=1,2,3,\dots$ by
$$Q(n)\equiv\sum_{k=1}^{\infty}\frac{n^{\underline k}}{n^{k}}=1+\frac{n-1}{n}+\frac{(n-1)(n-2)}{n^2}+\dots,$$
We have $\sum_{n=1}^{\infty}Q(n)n^{n-1}\frac{z^n}{n!}=-\log(1-T(z))$, where
$T(z)=\sum_{n=1}^{\infty}\frac{n^{n-1}}{n!}\,z^n=z\exp(T(z))$
is the egf for rooted labelled trees.
To get the large-$n$ asymptotics of $Q$, first consider the related
function [fgkp95]
$$R(n)\equiv 1+\frac{n}{n+1}+\frac{n^2}{(n+1)(n+2)}+\dots,\;n=1,2,3,\dots$$
and let $D(n)=R(n)-Q(n)$.  We may immediately deduce:
\begin{enumerate}
 \item $Q(n)+R(n)=n!\,e^n/n^n$
 \item $\sum_{n=1}^{\infty}D(n)n^{n-1}\frac{z^n}{n!}=\log[\frac{(1-T(z))^2}{2(1-ez)}]$
 \item $D(n)\sim \sum_{k=1}^{\infty}c(k)[z^n](T(z)-1)^k$, where 
 $c(k)\!\equiv\![\delta^k]\log(\delta^2/(2(1-(1+\delta)e^{-\delta})))$
 \item 
 $D(n)\sim 
 \frac{2}{3}+{\frac {8}{135}}\,{n}^{-1}-{\frac {16}{2835}}\,{n}^{-2}-{\frac {32
}{8505}}\,{n}^{-3}+{\frac {17984}{12629925}}\,{n}^{-4}+{\frac {668288}
{492567075}}\,{n}^{-5}+O \left( {n}^{-6} \right)
    $
\end{enumerate}
Now using $Q(n)=(n!\,e^n/n^n-D(n))/2$, we get the results shown in
table~\ref{resultsweget}.

\begin{table}
\begin{center}
\begin{tabular}{rrrrrrr}
\toprule
  & $[n^{1/2}]$ & $[n^0]$ & $[n^{-1/2}]$ & $[n^{-1}]$ & $[n^{-3/2}]$ & $[n^{-2}]$\\
\midrule
$D$ & 0 & $\frac{2}{3}$ & $\frac{8}{135}$ & $-\frac {16}{2835}$ & $-\frac{32}{8505}$ & $\frac{17984}{12629925}$\\
$Q$ & $\frac{\xi}{2}$ & $-\frac{1}{3}$ & $\frac{\xi}{24}$ & $-\frac{4}{35}$ & $\frac{\xi}{576}$ & $\frac{8}{235}$\\
\bottomrule
\end{tabular}
\end{center}
\caption{Coefficients in asymptotic expansions of $D$ and $Q$.}
\label{resultsweget}
\end{table}



\subsection{$W$}
Now let $W_{k}$ be the egf for connected labelled $(k+1)$-cyclic graphs.
It is known that [jklp93]:
  \begin{enumerate}
    \item for unrooted trees $W_{-1}(z)=T(z)-T^2(z)/2$, $[z^n]W_{-1}(z)=n^{n-2}$
    \item for unicycles $W_0(z)=-(\log(1-T(z))+T(z)+T^2(2)/2)/2=\textstyle{1\over 3!}z^3 +{15\over 4!}z^4+{222\over 5!}z^5+{3660\over 6!}z^6+\dots$
    \item for bicycles $W_{1}(z)=\frac{6T^4(z)-T^5(z)}{24(1-T(z))^3}=\textstyle{6\over 4!}z^4+{205\over 5!}z^5+{5700\over 6!}z^6+\dots$
    \item for $k\geqslant 1$, $W_{k}(z)=\frac{A_k(T(z))}{(1-T(z))^{3k}}$, where 
    $A_k$ are polynomials explicitly computable from results in [jklp93]
  \end{enumerate}

\subsection{$t$}
Knuth and Pittel's tree polynomials $t_n(y)$ ($y\neq 0$) are defined by
$$(1-T(z))^{-y}=\sum_{n=0}^{\infty}\,t_n(y)\frac{z^n}{n!}.$$
We can compute these for $y>0$ from the recurrence
\begin{eqnarray}
t_n(1)&=&1\nonumber\\
t_n(2)&=&n^n(1+Q(n))\nonumber\\
t_n(y+2)&=&(n/y)\,t_n(y)+t_n(y+1),\quad y>0\nonumber
\end{eqnarray}
Thanks to this recurrence, the asymptotics for each $t_n$ follows from the
known asymptotics of $Q$.  To apply these results to the problem of
asymptotically expanding $c(n,n+k)$, we need to express $c(n,n+k)$ as a linear
combination of values of $t_n(l)$ at integers $l$, which is always possible by
solving a linear system.  Some examples for small $k$ follow.
\begin{eqnarray}
c(n,n)   &=& n![z^n]W_0(z)\nonumber\\
         &=& \frac{1}{2}Q(n)n^{n-1}+3/2+t_n(-1)-t_n(-2)/4\nonumber\\
c(n,n+1) &=& n![z^n]W_1(z)\nonumber\\
         &=& \frac{5}{24}t_n(3)-\frac{19}{24}t_n(2)+\frac{13}{12}t_n(1)-\frac{7}{12}t_n(0)+\frac{1}{24}t_n(-1)+\frac{1}{24}t_n(-2)\nonumber
\end{eqnarray}

We finally have the desired exact results for $c(n,n+k)$ as shown in
table~\ref{exact-cnr}.  The conjectured numerical results of table~\ref{table1}
are confirmed and it appears that Flajolet \emph{et al}. are in error.
\begin{table}\begin{center}
\begin{tabular}{rrrrrrr}
\toprule
$k$ & $[n^0]$ & $[n^{-1/2}]$ & $[n^{-1}]$ & $[n^{-3/2}]$& $[n^{-2}]$ & $[n^{-5/2}]$\\
\midrule
$0$ & $\xi\frac{1}{4}$ & $-\frac{7}{6}$ & $\xi\frac{1}{48}$ & $\frac{131}{270}$ & $\xi\frac{1}{1152}$ & $-\frac{4}{2835}$\\
$1$ & $\frac{5}{24}$ & $-\xi\frac{7}{24}$ & $\frac{25}{36}$ &$-\xi\frac{7}{288}$ & $-\frac{79}{3240}$ & $-\xi\frac{7}{6912}$\\
$2$ & $\xi\frac{5}{256}$ & $-\frac{35}{144}$ & $\xi\frac{1559}{9216}$ & $-\frac{55}{144}$ & $\xi\frac{33055}{221184}$& $-\frac{41971}{136080}$ \\
\bottomrule
\end{tabular}
\end{center}
\caption{The asymptotic number of connected graphs: coefficients in the asymptotic expansion of $c(n,n+k)/n^{n+(3k-1)/2}$.}
\label{exact-cnr}
\end{table}


\section{Probability of connectivity} 
We now have all the results needed to calculate the asymptotic
probability $P(n,n+k)$ that a randomly chosen graph with $n$ nodes and $n+k$
edges is connected (for $n\rightarrow\infty$ and small fixed $k$).
The total number of graphs is $g(n,n+k)\equiv\binom{\binom{n}{2}}{n+k}$.
This can be asymptotically expanded for small $k$.   The results are in
table~\ref{tableg}.  Thus we get the final results for the probability of
connectivity in table~\ref{tablep}.

\begin{table}[th]
\begin{center}
\begin{tabular}{rrrrrrr}
\toprule
$k$ & $[n^0]$ & $[n^{-1}]$ & $[n^{-2}]$ & $[n^{-3}]$& $[n^{-4}]$ & $[n^{-5}]$\\
\midrule
$-1$ & $1$ & $\frac{7}{4}$ & $\frac{259}{96}$ & $\frac{22393}{5760}$ & $\frac{54359}{10240}$ & $\frac{52279961}{7741440}$\\
$ 0$ & $\frac{1}{2}$ & $-\frac{5}{8}$ & $-\frac{53}{192}$ & -$\frac{4067}{11520}$ & $-\frac{9817}{20480}$ & $-\frac{10813867}{15482880}$\\
$ 1$ & $\frac{1}{4}$ & -$\frac{21}{16}$ & $\frac{811}{384}$ & $-\frac{43187}{23040}$ & $\frac{159571}{73728}$ & $-\frac{55568731}{30965760}$\\
$ k$ & $\frac{1}{2^{k+1}}$ & ${\cal{O}}(1)$ & & & & \\
\bottomrule
\end{tabular}
\end{center}
\caption{The asymptotic total number of graphs: coefficients in the asymptotic expansion of $g(n,n+k)/(\sqrt{\frac{2}{\pi}}{e^{n-2}}{\left(\frac{n}{2}\right)}^{n}n^{(2k-1)/2})$.}
\label{tableg}
\end{table}
\begin{table}[h]
\begin{center}
\begin{tabular}{rrrrrr}
\toprule
$k$ & $[n^0]$ & $[n^{-1/2}]$ & $[n^{-1}]$ & $[n^{-3/2}]$ & $[n^{-2}]$\\
\midrule
$-1$ & $\frac{1}{2}$ & $0$ & $-\frac{7}{8}$ & $0$ & $\frac{35}{192}$\\ 
$0$ & $\frac{\xi}{4}$ & $-\frac{7}{6}$ & $-\frac{\xi}{3}$ & $-\frac{1051}{1080}$ & $\frac{5\xi}{9}$ \\
$1$ & $\frac{5}{12}$ & $\frac{-7\xi}{12}$ & $\frac{515}{144}$ & $-\frac{28\xi}{9}$ & $\frac{788347}{51840}$ \\
\bottomrule
\end{tabular}
\end{center}
\caption{The probability of connectivity: coefficients in the asymptotic expansion of $P(n,n+k)/(2^n e^{2-n} n^{k/2} \xi)$.}
\label{tablep}
\end{table}
\clearpage

\begin{figure}[thp]
\begin{center}
\includegraphics[width=0.7\hsize]{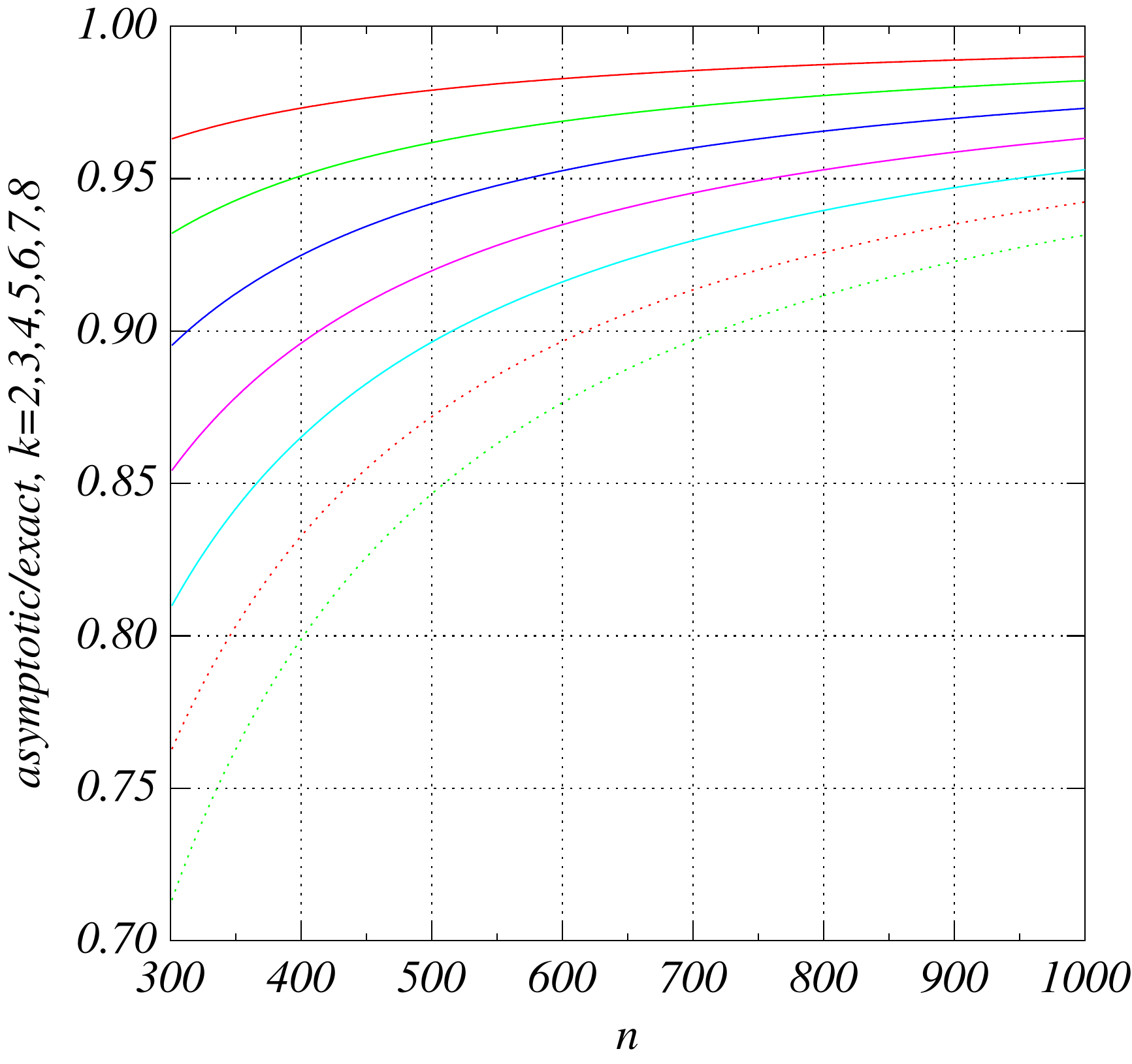}
\caption{Enumeration of labelled graphs - comparison of exact data with new asymptotic formulas.} 
\label{check}
\end{center}
\end{figure}

This method can easily be taken further further - it is simply a matter of symbol crunching.
In figure~\ref{check}, I confirm the accuracy of the new formulas by comparison with exact enumeration.
\clearpage

\section{References}
\begin{list1}
\item [{[ram11]}] S. Ramanujan: \emph{Question 294}, J. Indian Math. Soc. {\bf 3} 128 (1911) 
\item [{[bcm90]}] 
E. A. Bender, E. R. Canfield \& B. D. McKay: \emph{The asymptotic number of labeled connected graphs with a given number of vertices and edges},  Random structures and algorithms {\bf 1}, 127-169 (1990)
\item [{[slo03]}] N. J. A. Sloane, ed. (2003), The on-line encyclopedia of integer sequences\\
\url{www.research.att.com/~njas/sequences/}
\item [{[jklp93]}] S. Janson,  D. E. Knuth, T. Luczak \& B. G. Pittel: The birth of the giant component \emph{Random Structures and Algorithms},  {\bf 4}, 233-358 (1993)\\
\url{www-cs-faculty.stanford.edu/~knuth/papers/bgc.tex.gz}
\item [{[fgkp95]}] Ph Flajolet, P. Grabner, P. Kirschenhofer \& H. Prodinger: \emph{On Ramanujan's Q-function} Journal of Computational and Applied Mathematics
{\bf 58} 103-116 (1995)\\
\url{www.inria.fr/rrrt/rr-1760.html}
\item [{[fss04]}] Philippe Flajolet, Bruno Salvy and Gilles Schaeffer: \emph{Airy Phenomena and Analytic Combinatorics of Connected Graphs}\\
\url{www.combinatorics.org/Volume_11/Abstracts/v11i1r34.html}
\end{list1}
\end{document}